\documentclass[preprint]{llncs}

\usepackage{amsmath}
\usepackage{amssymb}
\usepackage{mathtools}
\usepackage{color}
\usepackage{algorithm}
\usepackage[noend]{algpseudocode}
\usepackage{listings}
\usepackage{multirow}
\usepackage{graphicx}
\usepackage{paralist}
\usepackage[title]{appendix}

\usepackage{xspace}
\newcommand{\toolname}{Z3str3\xspace}

\begin{document}
\title{\toolname{}: A String Solver with Theory-aware Branching}
\author{Murphy Berzish\inst{1} \and Yunhui Zheng\inst{2} \and Vijay Ganesh\inst{1}}
\institute{University of Waterloo, Waterloo, Canada
  \and
  IBM Research, Yorktown Heights, USA
  }

\maketitle

\begin{abstract}
We present a new string SMT solver, \toolname{}, that is faster than
its competitors Z3str2, Norn, CVC4, S3, and S3P over majority of three
industrial-strength benchmarks, namely, Kaluza, PISA, and IBM
AppScan. Z3str3 supports string equations, linear arithmetic over
length function, and regular language membership predicate. The key
algorithmic innovation behind the efficiency of Z3str3 is a technique
we call theory-aware branching, wherein we modify Z3's branching
heuristic to take into account the structure of theory literals to
compute branching activities. In the traditional DPLL(T) architecture,
the structure of theory literals is hidden from the DPLL(T) SAT solver
because of the Boolean abstraction constructed over the input theory
formula. By contrast, the theory-aware technique presented in this
paper exposes the structure of theory literals to the DPLL(T) SAT
solver's branching heuristic, thus enabling it to make much smarter
decisions during its search than otherwise. As a consequence,
\toolname{} has better performance than its competitors.
\end{abstract}

\section{Introduction}
\label{sec:intro}
String SMT solvers are increasingly becoming important for security
applications and in the context of analysis of string-intensive
programs~\cite{tacas09,emmiMS2007,hampi,willem,prateek,jalangi,WassermannSu2007}. Many
string SMT solvers, such as Z3str2~\cite{Z3str2-FMSD,Z3str2-CAV15}
(and its predecessor Z3str~\cite{z3-str}), CVC4~\cite{CVC4-CAV14},
Norn~\cite{norn}, S3~\cite{s3} (and its successor S3P~\cite{S3P}), and
Stranger (and its successor ABC~\cite{abc-cav15}) have been developed
to address these challenges and applications. We have developed
the \toolname{} string solver as a native first-class theory solver
directly integrated into the Z3 SMT solver~\cite{z3}, that is much
faster than its predecessors Z3str2 and Z3str, as well as competitors
CVC4, Norn, and S3.  Having direct access to the core solver of Z3 has
allowed us to develop and implement a novel DPLL(T) technique which we
call {\it theory-aware branching}, described below. We follow the
latest string SMT language standard supported by all major string
solvers, and published on the CVC4 website~\cite{CVC4-CAV14}.

\subsection{Contributions}

\begin{enumerate}

\item {\bf Theory-aware branching:} We leverage the integration between
the Z3 SMT solver's DPLL(T) SAT layer (henceforth referred to as the
core solver) and the string solver to guide the search and prioritize
certain branches of the search tree over others. In particular, we
modify the activity computations of the branching heuristic of the Z3
core solver, making the heuristic aware of the structure of the theory
literals underlying the Boolean abstraction of the input formula such
that ``simpler'' theory literals are prioritized over more complex
ones.

\item {\bf Theory-aware case-split:}
We add an optimization to Z3's core solver that enables efficient
representation of mutually exclusive Boolean variables in the Boolean
abstraction of the input theory formula.

\item {\bf Experimental evaluation:}
To validate the effectiveness of our techniques, we present a
comprehensive and thorough evaluation of \toolname{}, and compare
against Z3str2, CVC4, S3, and Norn on several large
industrial-strength benchmarks. We couldn't directly compare against
S3P since its source is not available, but do summarize the results
from their CAV 2016 paper and compare against \toolname{}. We also
couldn't compare against Stranger/ABC because these tools don't
produce models for SAT cases, don't support dis-equations over
arbitrary string terms, and have incorrectness issues as noted in
their own paper~\cite{abc-cav15}.
\end{enumerate}

\section{Theory-Aware Branching}
\label{sec:branching}

Several of the key enhancements we make in \toolname{} over Z3str2
involve changes to the core DPLL(T) SAT solver in Z3, which handles
the Boolean structure of the formula and performs propagation and
branching.  The first of these enhancements is referred to
as \textbf{theory-aware branching}.  We modify the Z3 core solver to
allow theory solvers to provide information about certain theory
literals that are given increased or decreased priority during the
search. For example, consider the case where the solver learns the
equality $X \cdot Y = A \cdot B$ for non-constant terms $X, Y, A,
B$. The behaviour of \toolname{} (in line with Z3str2) is to handle
this equality by considering a disjunction of three possible
arrangements~\cite{Z3str2-FMSD,Z3str2-CAV15}:

\begin{itemize}
\item $X = A$ and $Y = B$
\item $X = A \cdot s_1$ and $s_1 \cdot Y = B$ for a fresh non-empty string variable $s_1$
\item $X \cdot s_2 = A$ and $Y = s_2 \cdot B$ for a fresh non-empty string variable $s_2$
\end{itemize}

Of the three possible arrangements, the first is the simplest to check
because it does not introduce any new variables and only asserts
equalities between existing terms. Therefore, we would like Z3's core
solver to prioritize checking this arrangement before the others. The
advantage gained by theory-aware branching is the ability to give the
core solver information regarding the relative importance of each
branch, allowing the theory solver to exert additional control over
the search. Note we always prioritize simpler branches over more
complex ones.

We implement theory-aware branching as a modification of the branching
heuristic in Z3. This idea of creating a theory-aware DPLL branching heuristic
is mentioned in~\cite{LazySMT}.
The default branching heuristic in Z3 is
activity-based, similar to VSIDS~\cite{chaff}.  The core solver will
branch on the literal with the highest activity that has not yet been
assigned.  In the VSIDS branching heuristic, activity is increased
additively when a literal appears in a conflict clause, and decayed
multiplicatively at regular intervals.

The theory-aware branching technique computes the activity of a
literal $A$ as the sum of two terms $A_{b}$ and $A_{t}$, wherein the
term $A_{b}$ is the ``base activity'', which is the standard activity
of the literal as computed and handled by Z3's core solver. The term
$A_{t}$ is the ``theory-aware activity''.  The value of this term is
provided for individual literals by theory solvers, and is taken to be
0 if no theory-aware branching information has been provided.  This
modification causes the core solver to branch on the literal with the
highest activity $A$, taking into account both the standard activity
value and the theory-aware activity.  Therefore, assigning a (small)
positive theory-aware activity to a literal will cause it to have
higher activity than usual, making it more likely for the core solver
to choose it to branch on.  Conversely, assigning a (small) negative
theory-aware activity will deter the core solver from choosing that
literal. Theory-aware branching in Z3str3 modifies the activities of
theory literals as follows:

\begin{enumerate}

\item Literals corresponding to arrangements that do not create new variables are given a large (0.5)
$A_t$. Other arrangements are given a small (0.1) $A_t$.

\item Arrangements that allow a variable to become equal to a constant
string are given a small (0.2) $A_t$.

\item When searching for length of strings, we give the literal
corresponding to the choice ``generate more length options'' a small negative (-0.1) $A_t$.

\end{enumerate}

\section{Theory-Aware Case-split}
\label{sec:theorycasesplit}

During the search, a theory solver can create terms which encode a
disjunction of Boolean literals that are pairwise mutually exclusive,
i.e.  exactly one of the literals must be assigned true and the others
must be assigned false.  We refer to this as a \textbf{theory-aware
case-split}. As an example, consider the case where the string solver
learns that a concatenation of two arbitrary string terms $X$ and $Y$
is equal to a string constant $c = c_1 c_2 \hdots c_n$ of length $n$,
where each $c_i$ is a character in $c$.  At this point there are $n+1$
possible ways in which we can split the constant $c$ over $X$ and $Y$
resulting in different arrangements:

\begin{itemize}
\item $X = \epsilon, Y = c_1 c_2 \hdots c_n$
\item $X = c_1, Y = c_2 c_3 \hdots c_n$
\item $\hdots$
\item $X = c_1 c_2 \hdots c_n, Y = \epsilon$
\end{itemize}

Note that each of these arrangements represents a case that can be
explored by the solver, and also that all of these cases are mutually
exclusive (as clearly $X$ cannot be equal to both $\epsilon$ and $c_1$
simultaneously, etc.).  Thus, this represents a theory-aware
case-split. Note that the Boolean abstraction constructed over theory
literals completely hides this obviously useful information that these
variables (and corresponding arrangements) are mutually exclusive. A
naive solution is to encode $O(n^2)$ extra mutual exclusion Boolean
clauses over these variables. Unfortunately, this would result in very
poor performance because of the quadratic blowup in formula
size. Another option is to let the congruence closure solver in the
Z3 core discover the mutual exclusivity of these Boolean
variables. This can result in unnecessary backtracking, unnecessary
calls to congruence closure, and, in the worst case, reduces to the
same set of mutual exclusion clauses being learned in the form of
conflict clauses. We improve the performance of theory case-splits by
allowing theory solvers to provide extra information to the core
solver regarding which literals can be treated as mutually exclusive
during its search.  This means that theory solvers do not have to
assert extra clauses to enforce mutual exclusivity of
choices. Instead, our modified core solver can now use this extra
information from the theory solver during its search. Our
implementation of this technique is as follows:

\begin{enumerate}
  \item The theory solver provides the core solver with a set $S$ of
  mutually exclusive literals that correspond to a theory
  case-split. This set is maintained by the core solver in a list of
  all such sets.

  \item During branching, the core solver checks if the current
  branching literal belongs to some such set $S$. If yes,
  the current branching literal is assigned true and all other
  theory case-split literals in $S$ are assigned false.
  Otherwise, the default branching behaviour is used.

  \item During propagation, the core solver may assign a truth value
  to a literal $l$ in some set $S$ of theory case-split literals.  If
  so, the theory case-split check is invoked, i.e., the core solver
  checks whether two literals $l_1, l_2$ in the same set $S$ have been
  assigned the value true. If this is the case, the core solver
  immediately generates the conflict clause $(\lnot l_1 \lor \lnot
  l_2)$.
\end{enumerate}

\section{Experimental Results}
\label{sec:results}
\begin{figure*}[t!]
\centering
\begin{tabular}{cc}
\includegraphics[width=0.5\textwidth]{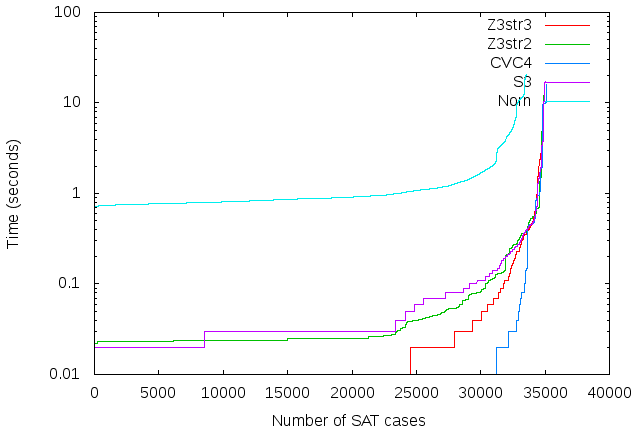} & \includegraphics[width=0.5\textwidth]{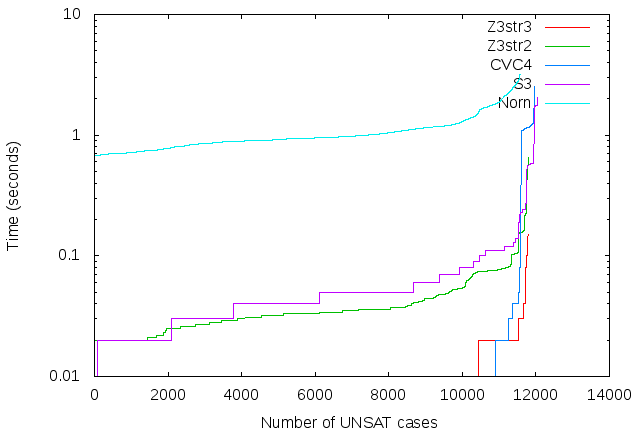} \\[\abovecaptionskip]
\small (a) SAT cases & \small (b) UNSAT cases
\end{tabular}
\caption{Cactus plots for the Kaluza benchmark suite.}
\label{fig:kaluza}
\end{figure*}

In this section, we describe the experimental evaluation of
the \toolname{} solver to validate the effectiveness of the techniques
presented in this paper. All techniques improve solver efficiency in
isolation as well as in combination. In the interest of brevity we
only report on the combined result in detail.  We compare \toolname{}
against four other state-of-the-art string solvers, namely,
Z3str2~\cite{Z3str2-CAV15,Z3str2-FMSD}, CVC4~\cite{CVC4-CAV14},
S3~\cite{s3}, and Norn~\cite{norn}, across industrial benchmarks
obtained from Kaluza~\cite{kaluza}, PISA~\cite{PISA}, and AppScan
Source~\cite{AppScan}. Each of these benchmark suites draw from
real-world applications with diverse characteristics. All experiments
were performed on a workstation running Ubuntu 15.10 with an Intel
i7-3770k CPU and 16GB of memory. Also, we cross-verified the models
generated by \toolname{} against Z3str2 and CVC4, and vice-versa.


Table~\ref{table:kaluza} shows the summary of results for the Kaluza
benchmark. Figure~\ref{fig:kaluza} presents the results in two cactus
plots. Binaries for S3P are not publicly available so we were not able
to evaluate it directly or include the timing information in the
cactus plots in Figure~\ref{fig:kaluza}.  Instead, we report the
aggregate results presented for this benchmark in the most recent S3P
paper~\cite{S3P}.  From Table~\ref{table:kaluza} it is clear
that \toolname{} is competitive with respect to CVC4, and is much
faster than other tools. CVC4 has 193 errors, but \toolname{} times out
on 84 cases. The unknowns in \toolname{} are because it lacks the
feature to handle string equations with overlapping variables, similar
to Z3str2.

\begin{table}[t!]
\centering
\begin{tabular}[t]{|c|c|c|c|c|c|c|}
\hline
& Z3str3 & Z3str2 & CVC4 & Norn & S3 & S3P \\
\hline
\hline
sat & 34885 & 34868 & 35128 & 33527 & 35016 & 35270 \\
unsat & 11786 & 11799 & 11957 & 11568 & 12049 & 12014 \\
\hline
\hline
unknown & 529 & 617 & 6 & 1913 & 0 & 0\\
timeout & 84 & 0 & 0 & 276 & 219 & 0\\
error & 0 & 0 & 193 & 0 & 0 & 0\\
crash & 0 & 0 & 0 & 0 & 0 & 0\\
\hline
\hline
Total time (s) & 5275.11 & 3997.63 & 4851.66 & 109280.76 & 10544.06 & 6972 \\
\hline
Total time without timeouts (s) & 3595.11 & 3997.63 & 4851.66 & 97784.00 & 6164.06 & 6972 \\
\hline
\end{tabular}
\vspace{10pt}
\caption{Results on cases from the Kaluza benchmark. Timeout=20~s.}
\label{table:kaluza}
\end{table}

Table~\ref{table:pisa} shows the results on the PISA benchmark. Norn
was not able to solve any of the cases as it crashed upon seeing
unrecognized string operators (e.g. {\tt indexof}).  From
Table~\ref{table:pisa} we make the following observations.  The
tools \toolname{}, Z3str2, and CVC4 are in agreement on all cases they
are able to solve, with CVC4 and Z3str2 timing out on one SAT case
which \toolname{} can solve in 16.58 seconds. The results for S3 are
significantly worse; it is unable to solve {\tt pisa-009.smt2} while
the other three solvers all answer SAT reasonably quickly, and in
addition S3 incorrectly answers UNSAT for {\tt pisa-008.smt2}, {\tt
pisa-010.smt2}, and {\tt pisa-011.smt2}, on which \toolname{} and (for
two of these cases) Z3str2 and CVC4 all return SAT and produce a valid
model.

\begin{table}[t!]
\centering
\begin{tabular}[t]{|c|c|c|c|c|c|c|c|c|}
\hline
\multirow{2}{*}{input} & \multicolumn{2}{|c|}{\toolname{}} & \multicolumn{2}{|c|}{Z3str2} & \multicolumn{2}{|c|}{CVC4} & \multicolumn{2}{|c|}{S3} \\
\cline{2-9}
& result & time (s) & result & time (s) & result & time (s) & result & time (s) \\
\hline
pisa-000.smt2 & sat & 0.03 & sat & 0.25 & sat & 0.08 & sat & 0.07 \\
\hline 
pisa-001.smt2 & sat & 0.01 & sat & 0.19 & sat & 0.00 & sat & 0.07 \\
\hline
pisa-002.smt2 & sat & 0.01 & sat & 0.10 & sat & 0.00 & sat & 0.05 \\
\hline
pisa-003.smt2 & unsat & 0.00 & unsat & 0.02 & unsat & 0.01 & unsat & 0.02 \\
\hline
pisa-004.smt2 & unsat & 0.01 & unsat & 0.05 & unsat & 0.39 & unsat & 0.05 \\
\hline
pisa-005.smt2 & sat & 0.06 & sat & 0.14 & sat & 0.02 & sat & 0.04 \\
\hline
pisa-006.smt2 & unsat & 0.01 & unsat & 0.05 & unsat & 0.32 & unsat & 0.05 \\
\hline
pisa-007.smt2 & unsat & 0.01 & unsat & 0.05 & unsat & 0.37 & unsat & 0.05 \\
\hline
pisa-008.smt2 & sat & 16.58 & timeout & 20.00 & timeout & 20.00 & unsat \text{\sffamily X} & 4.73 \\
\hline
pisa-009.smt2 & sat & 12.59 & sat & 0.62 & sat & 0.00 & timeout & 20.00 \\
\hline
pisa-010.smt2 & sat & 0.03 & sat & 0.09 & sat & 0.00 & unsat \text{\sffamily X} & 0.02 \\
\hline
pisa-011.smt2 & sat & 0.04 & sat & 0.06 & sat & 0.00 & unsat \text{\sffamily X} & 0.02 \\
\hline
\end{tabular}
\vspace{10pt}
\caption{PISA benchmark results. Timeout=20~s. \text{\sffamily X} = incorrect response.}
\label{table:pisa}
\end{table}

\begin{table}[t!]
\centering
\begin{tabular}[t]{|c|c|c|c|c|c|c|c|c|}
\hline
\multirow{2}{*}{input} & \multicolumn{2}{|c|}{\toolname{}} & \multicolumn{2}{|c|}{Z3str2} & \multicolumn{2}{|c|}{CVC4} & \multicolumn{2}{|c|}{S3} \\
\cline{2-9}
& result & time (s) & result & time (s) & result & time (s) & result &
time (s) \\
\hline
t01.smt2 & sat & 7.05 & sat & 1.31 & sat & 0.01 & sat & 0.23 \\
\hline
t02.smt2 & sat & 0.13 & sat & 0.38 & sat & 0.01 & unknown & 0.04 \\
\hline
t03.smt2 & sat & 0.53 & sat & 9.54 & sat & 3.82 & sat \text{\sffamily X} & 0.14 \\
\hline
t04.smt2 & sat & 0.68 & sat & 4.45 & timeout & 20.00 & sat \text{\sffamily X} & 0.10 \\
\hline
t05.smt2 & sat & 1.15 & sat & 16.84 & sat & 3.87 & sat \text{\sffamily X} & 0.55 \\
\hline
t06.smt2 & sat & 0.02 & sat & 0.15 & sat & 0.01 & sat & 0.13 \\
\hline
t07.smt2 & sat & 2.62 & sat & 0.25 & sat & 0.00 & unknown & 0.02 \\
\hline
t08.smt2 & sat & 0.01 & sat & 0.25 & sat & 0.17 & sat \text{\sffamily X} & 0.03 \\
\hline
\end{tabular}
\vspace{10pt}
\caption{AppScan benchmark results. Timeout=20~s. \text{\sffamily X} = incorrect response.}
\label{table:appscan}
\end{table}

Table~\ref{table:appscan} shows the results on the AppScan
benchmark. Norn crashed on these cases as well upon seeing
unrecognized string operators. From Table~\ref{table:appscan} we make
the following observations. \toolname{}, Z3str2, and CVC4 all agree
on all cases they are able to solve. CVC4 performs better
than \toolname{} on 3 cases and worse on 5 (including one timeout).
Z3str2 performs better than \toolname{} on 2 cases and worse on 6,
taking almost three times as long in total (33.17 seconds vs. 12.19
seconds).  S3 returns UNKNOWN on two cases that are
solved by the other three tools and produces invalid models
which fail cross-validation for four other cases.

\section{Discussion on Experimental Results, and Conclusions}

The experimental results discussed here make clear the efficacy of
theory-aware branching and case-split. The crucial insight behind
these techniques is that biasing the search towards easier branches of
the search tree (e.g., an arrangement that doesn't require splitting
variables, as opposed to one with overlapping variables) is often very
effective since most string constraints obtained from practical
applications have the ``small model'' property. The slogan of
theory-aware branching is ``bias search towards easy cases first''. We
also note that \toolname{} and CVC4 are sound, and more robust as
compared to Norn and S3 which sometimes give wrong answers or crash on
the benchmarks we used.

\bibliographystyle{abbrv}
\bibliography{main,cav15}
\end{document}